\documentclass{article}
\usepackage{amsmath,amssymb,amsfonts,amsthm}
\usepackage[russian]{babel}
\usepackage{color}
\usepackage{graphicx}
\usepackage{wrapfig}

\begin{document}

\flushbottom

\renewcommand{\figurename}{Fig.}
\def\refname{References}
\def\proofname{Proof}

\newtheorem{teo}{Theorem}
\newtheorem{pro}{Proposition}
\newtheorem{rem}{Remark}

\def\tens#1{\ensuremath{\mathsf{#1}}}

\if@mathematic
   \def\vec#1{\ensuremath{\mathchoice
                     {\mbox{\boldmath$\displaystyle\mathbf{#1}$}}
                     {\mbox{\boldmath$\textstyle\mathbf{#1}$}}
                     {\mbox{\boldmath$\scriptstyle\mathbf{#1}$}}
                     {\mbox{\boldmath$\scriptscriptstyle\mathbf{#1}$}}}}
\else
   \def\vec#1{\ensuremath{\mathchoice
                     {\mbox{\boldmath$\displaystyle\mathbf{#1}$}}
                     {\mbox{\boldmath$\textstyle\mathbf{#1}$}}
                     {\mbox{\boldmath$\scriptstyle\mathbf{#1}$}}
                     {\mbox{\boldmath$\scriptscriptstyle\mathbf{#1}$}}}}
\fi

\begin{center}
{\Large\bf Hamiltonization of Elementary Nonholonomic~Systems\\}

\bigskip

\footnotetext{The work of I.\,A.\,Bizyaev was carried out within the
framework of the basic part of the State assignment to institutions of
higher education. The work of A.\,V.\,Borisov was supported by the Russian
Foundation for Basic Research under grant no.~13-01-12462-ofi\_m and by
a~grant of the President of Russian Federation ``Supporting Leading
Scientific Schools'' NSh-2964.2014.1. Sections~3 and~4 were prepared by
I.\,S.\,Mamaev within the framework of the grant of Russian Science
Foundation (project no.~14-19-01303).}

{\large\bf Ivan~A.\,Bizyaev$^1$,
Alexey~V.\,Borisov$^2$,
Ivan~S.\,Mamaev$^3$\\}
\end{center}

\begin{quote}
\begin{small}
\noindent
$^1$ Udmurt State University, Universitetskaya 1, Izhevsk, 426034,
Russia\\
$^2$ Steklov Mathematical Institute of Russian Academy of Sciences, Gubkina
ul.~8, 119991 Moscow, Russia,\\
Udmurt State University, Universitetskaya ul. 1, 426034 Izhevsk, Russia\\
Russia. E-mail: borisov@rcd.ru\\
$^3$ Izhevsk State Technical University, Studencheskaya ul. 7, 426069 Izhevsk,
Russia
\end{small}

\bigskip
\bigskip

\begin{small}
\textbf{Abstract.}  In this paper, we develop the Chaplygin reducing
multiplier method; using this method, we obtain a conformally
Hamiltonian\linebreak representation for three nonholonomic systems,
namely, for the\linebreak nonholonomic oscillator, for the Heisenberg
system, and for the Chaplygin sleigh. Furthermore, in the case of an
oscillator and the nonholonomic Chaplygin sleigh, we show that the problem
reduces to the study of motion of a mass point (in a potential field) on a
plane and, in the case of the Heisenberg system, on the sphere. Moreover,
we consider an example of a~nonholonomic system (suggested by Blackall) to
which one cannot apply the reducing multiplier method.

\smallskip


\smallskip

\end{small}
\end{quote}

\clearpage

\section{Introduction}

In the present paper, a few fairly simple (model) dynamical systems
of\linebreak nonholonomic mechanics are considered in connection with the
Hamiltonization problem, that is, the problem of reducing these systems to
Hamiltonian form. For meaningful problems of the theory of nonholonomic
systems (describing, as a rule, the dynamics of systems with rolling), see
the surveys \cite{13, 17}, and for the general Hamiltonization problem (of
general dynamical systems), see~\cite{7}.

The paper discusses the problem of motion of a mass point on a
three-dimensional Riemannian manifold in the presence of a
nonholonomic\linebreak (nonintegrable) constraint and a potential field.
If the constraint is integrable and the potential is absent, then the
problem is reduced to the well-studied case of motion along geodesics on a
two-dimensional manifold. As is well known, in this case, the behavior~of
geodesics is related to a variational problem, and its global aspect is
related to methods of the calculus of variations in general. The case of
nonzero potential can also be reduced to the problem of geodesics,
however, with a different metric, namely, the Maupertuis metric~\cite{40}.
In this connection we mention the work~\cite{30a}, in which an example is
given of the geodesic flow with integrals rational in the velocities.

In general, the nonintegrability of a constraint is incompatible with the
variational principle, which was already clear to Hertz, Poincar\'{e} and
Hamel~\cite{24}. Nevertheless, in some cases, by a suitable
reparameterization of time (depending on the configuration variables
only), the trajectories of the system can again be obtained using the
Hamilton variational principle, and the equations of motion are
represented in a conformally Hamiltonian form. The most natural methods of
Hamiltonization can be developed for Chaplygin systems by using his
reducing multiplier theory.

In this paper, we discuss three model problems (the nonholonomic
oscillator, the Heisenberg system, and the Chaplygin sleigh) for which the
Hamiltonization can be carried out explicitly and, moreover, we consider
another system (which was introduced by Blackall) for which there are
essential obstacles to the\linebreak Hamiltonization (in the entire phase
space) and, as a result of these obstacles, the behavior of the system
differs significantly from that for the variational problem of geodesics.

\section{Equations of motion}

Consider a (mechanical) system with three degrees of freedom and generalized
coordinates $q_1$, $q_2$, $q_3$.

Suppose that the coordinate $q_3$ is cyclic, i.e., that it is not included
explicitly in the Lagrangian of the system $L$. Moreover, we assume that
one can always carry out a Legendre transform of the Lagrangian~$L$.

Assume that the motion is subject to a nonholonomic constraint which is
linear, homogeneous in the velocities, and can be represented as
\begin{equation}
\label{eq_1}
f=\dot q_3-a_1(\bold q)\dot q_1-a_2(\bold q)\dot q_2=0, \quad \bold
q=(q_1,q_2).
\end{equation}

Let us write the equations of motion in the form of the Euler--Lagrange
equations with the undetermined multiplier $\lambda$,
\begin{equation}
\label{eq_2}
\frac d{dt}\left(\frac{\partial L}{\partial\dot{q_i}}\right)-\frac{\partial
L}{\partial q_i}=\lambda\frac{\partial f}{\partial\dot{q_i}}, \qquad i=1,2,3.
\end{equation}
It follows from the last equation that
$$
\lambda=\frac d{dt}\left(\frac{\partial L}{\partial\dot{q_3}}\right)\!\!.
$$

Denote by $L^*(\bold q,\dot{\bold q})$ the Lagrangian of the system after
substituting into it the expression for $\dot q_3$ from the constraint
equation. Using the standard rule of indirect differentiation, we see that
$$
\frac{\partial L^*}{\partial\dot q_i}=\frac{\partial L}{\partial\dot
q_i}+\frac{\partial L}{\partial\dot q_3}a_i, \qquad \frac{\partial
L^*}{\partial q_i}=\frac{\partial L}{\partial q_i}+\sum_{k=1}^2\frac{\partial
L}{\partial\dot q_3} \frac{\partial a_k}{\partial q_i}\dot q_k, \qquad i=1,2.
$$
Substituting the last relations into equations~\eqref{eq_2} and reducing similar
terms, we obtain a close system for the variables $(\bold q,\dot{\bold q})$,
\begin{equation}
\label{eq_3}
\begin{gathered}
\frac d{dt}\left(\frac{\partial
L^*}{\partial\dot{q_1}}\right)-\frac{\partial L^*}{\partial q_1}=S\dot{q_2},
\quad
\frac d{dt}\left(\frac{\partial L^*}{\partial\dot{q_2}}\right)
-\frac{\partial L^*}{\partial q_2}=-S\dot{q_1},\\
S=\left(\frac{\partial L}{\partial\dot q_3}\right)^*\left(\frac{\partial
a_1}{\partial q_2}- \frac{\partial a_2}{\partial q_1}\right),
\end{gathered}
\end{equation}
where the expression $\left(\frac{\partial L}{\partial\dot q_3}\right)^*$ means
that the substitution $\dot q_3$ is made after the differentiation.

Thus, the problem reduces to the study of the system (3) with two degrees
of freedom; according to \cite{15}, we refer to this system as the
generalized Chaplygin system. For known solutions $q_1(t)$ and $q_2(t)$,
the law of modification of the remaining variable $q_3$ is obtained,
according to~\eqref{eq_1}, by a quadrature. Further, we concentrate on the
investigation of system~\eqref{eq_3}. To study the integrability of systems
of this kind, one can use the results of \cite{29, 30, 37}.

\begin{rem}
The nonholonomicity of a constraint means that this constraint
cannot be represented in the form
$$
F(q_1,q_2,q_3)=0, \quad \text{where} \quad \frac{dF}{dt}=f,
$$
which implies the condition
\begin{equation}
\label{eq_4}
\frac{\partial a_1}{\partial q_2}\ne\frac{\partial a_2}{\partial q_1},
\end{equation}
which is assumed to be valid everywhere below.
\end{rem}

\section{Invariant measure and the reducing multiplier method}

As is well known (see, e.g., \cite{9, 28}), the equations of motion in
nonholonomic mechanics generally cannot be represented in Hamiltonian
form. Nevertheless, there are problems in which such a representation can
be obtained only after rescaling time, i.e., the equations of motion are
represented in conformally Hamiltonian form.

The {\it reducing multiplier\/} method is the most efficient to reduce the
generalized Chaplygin systems to conformally Hamiltonian form, and we
proceed with a~presentation of the method (see also \cite{8,
15}).

First of all, note that the homogeneity in the generalized velocities of
the constraint \eqref{eq_1} results in the fact that the system \eqref{eq_3}
preserves the energy integral
\begin{equation}
\label{eq_5}
E=\sum_{i=1}^2\dfrac{\partial L^*}{\partial\dot q_i}\dot q_i-L^*.
\end{equation}
Making the Legendre transform
$$
P_i=\frac{\partial L^*}{\partial\dot q_i}, \quad H=\sum_{i=1}^2P_i\dot
q_i-L^*\Bigr|_{\dot q_i\to P_i}, \qquad i=(1,2),
$$
we obtain the following system of equations:
\begin{equation}
\label{eq_6}
\dot q_i=\frac{\partial H}{\partial P_i}, \qquad \dot P_1=-\frac{\partial
H}{\partial q_1}+\frac{\partial H}{\partial P_2}S, \qquad \dot
P_2=-\frac{\partial H}{\partial q_2}-\frac{\partial H}{\partial P_1}S.
\end{equation}
Here $H$ stands for the integral \eqref{eq_5}, which, together with $S$, is
expressed in terms of the new variables.

{\bf Invariant measure.} Let us find cases in which system~\eqref{eq_6} has
an invariant measure that can be represented as
\begin{equation}
\label{eq_7}
\mathcal{N}(\bold q)d\bold qd\bold P.
\end{equation}

Recall that the function $\mathcal{N}(\bold q)$ is called the density of the
invariant measure and satisfies the Liouville equation~\cite{42, 22}
\begin{equation}
\label{eq_8}
\operatorname{div}(\mathcal{N}\bold v)=0,
\end{equation}
where $\bold v$ is the vector field determined by~\eqref{eq_6}.

\begin{rem}
As a rule, it is assumed that the density of the invariant
measure is a smooth and positive function on the entire phase space.
Nevertheless, in applications, one can face a situation in which $\mathcal{N}(\bold q)$ has singularities in some domain of the phase space. In this
case, the system is said to admit a {\it singular\/} invariant measure.
\end{rem}

Consider in more detail the Liouville equation \eqref{eq_8}, which in this
case can be represented as
\begin{equation}
\label{eq_9}
\left(\frac\partial{\partial\bold q}\ln\mathcal{N}(\bold q)+\bold\xi,\dot{\bold
q}\right)=0, \qquad \bold\xi=\left(-\frac{\partial S}{\partial
P_2},\frac{\partial S}{\partial P_1}\right).
\end{equation}
Since the previous relation must hold for arbitrary $\dot{\bold q}$, it follows
that
$$
\frac\partial{\partial\bold q}\ln\mathcal{N}(\bold q)+\bold\xi=0.
$$
In this case, for the solution $\mathcal{N}(\bold q)$ of \eqref{eq_9} to exist,
it is necessary that the vector field $\bold\xi$ be potential. This
condition leads to the relation
\begin{equation}
\label{eq_10}
\frac{\partial^2S}{\partial q_1\partial P_1}+\frac{\partial^2S}{\partial
q_2\partial P_2}=0.
\end{equation}
Thus, the following proposition holds.

\begin{pro}
If system \eqref{eq_6} has an invariant measure
\eqref{eq_7}, then \eqref{eq_10} holds.
\end{pro}

Note that, for systems for which the nonholonomic model admits an
invariant measure, one should take into account the friction forces to
describe the\linebreak asymptotic behavior~\cite{36}.

{\bf Reducing multiplier method.} Suppose we are given an invariant
measure $\mathcal{N}(\bold q)$. Then we make the following change of variables
$$
P_i=\frac{p_i}{\mathcal{N}(\bold q)}, \qquad i=1,2.
$$
Denote the functions in the new variables by $\overline H(\bold q,\bold
p)=H(\bold q,\bold P(\bold q,\bold p))$ and $\overline S(\bold q,\bold
p)=S(\bold q,\bold P(\bold q,\bold p))$, respectively. Then the following
relations hold for the derivatives:
$$
\begin{aligned}
\frac{\partial H}{\partial P_i}&=\mathcal{N}\frac{\partial \overline
H}{\partial p_i}, \quad \frac{\partial S}{\partial P_i}=\mathcal{N}\frac{\partial
\overline S}{\partial p_i},\\
\frac{\partial H}{\partial q_i}&=\frac{\partial\overline H}{\partial
q_i}+\frac1{\mathcal{N}}\frac{\partial \mathcal{N}}{\partial q_i}\left(\frac{\partial
\overline H}{\partial p_1}p_1+\frac{\partial \overline H}{\partial
p_2}p_2\right)\!.
\end{aligned}
$$
Further, substituting these relations into~\eqref{eq_6} and using~\eqref{eq_9}, we
obtain
\begin{equation}
\label{eq_11}
\begin{gathered}
\dot q_i=\mathcal{N}\frac{\partial\overline H}{\partial p_i}, \qquad \dot
p_1=\mathcal{N}\left(-\frac{\partial\overline H}{\partial q_1}+
K\frac{\partial\overline H}{\partial p_2}\right)\!, \qquad \dot p_2=\mathcal{N}\left(-\frac{\partial\overline H}{\partial q_2}-
K\frac{\partial\overline H}{\partial p_1}\right)\!,\\
K=\mathcal{N}\left(\overline S-\frac{\partial\overline S}{\partial
p_1}p_1-\frac{\partial\overline S}{\partial p_2}p_2\right)\!.
\end{gathered}
\end{equation}
Write $\bold x=(\bold p,\bold q)$ and represent the system~\eqref{eq_11} as
$$
\dot{\bold x}=\mathcal{N}{\text\bf J}\frac{\partial\overline H}{\partial\bold
x},\qquad {\text\bf J}=
\begin{pmatrix}
0&K&1&0\\
-K&0&0&1\\
-1&0&0&0\\
0&-1&0&0
\end{pmatrix}\!.
$$
For the skew-symmetric matrix ${\text\bf J}$ to define the Poisson bracket
$$
\{q_i,q_j\}=0, \qquad \{q_i,p_j\}=\delta_{ij}, \qquad \{p_1,p_2\}=K, \qquad
i=1,2,
$$
and, thus, for the invariant measure $\mathcal{N}$ to be a reducing multiplier,
it is necessary that the Jacobi identity be valid, which in this case is
reduced to the following two equations for $\overline S$:
\begin{equation}
\label{eq_12}
p_1\frac{\partial^2\overline S}{\partial p_1^2}+p_2\frac{\partial^2\overline
S}{\partial p_1\partial p_2}=0, \quad p_2\frac{\partial^2\overline S}{\partial
p_2^2}+p_1\frac{\partial^2\overline S}{\partial p_1\partial p_2}=0.
\end{equation}

If $\overline S$ is a linear function in $p_1$ and $p_2$, then the
previous relation holds, and thus the equations of motion \eqref{eq_11} can
be represented in conformally Hamiltonian form.

Obviously, the linearity of $\overline S$ means that $\dot q_1$ and $\dot
q_2$ enter the Lagrangian $L^*$ linearly and quadratically, which, as a
rule, occurs in practice. We formulate the result thus obtained more
clearly in the form of the following theorem.

\begin{teo}
{\rm (reducing multiplier method)} If a system with a
constraint of the form~\eqref{eq_1} and the Lagrangian
\begin{equation}
\label{eq_13}
L=\frac12\sum_{i,j=1}^2g_{ij}(\bold q)\dot q_i\dot q_j+\sum_{i=1}^2c_i(\bold
q)\dot q_i-U(\bold q),
\end{equation}
where $\bold q=(q_1,q_2)$ and $\bold g=\|g_{ij}(\bold q)\|$ is a symmetric
matrix, admits a smooth invariant measure with density $\mathcal{N}(\bold
q),$ then the equations of motion {\rm (}on the entire phase space{\rm )}
can be represented in conformally Hamiltonian form.
\end{teo}

\proof Indeed, if \eqref{eq_10} holds, then, using the solution of the
Liouville equation, one can represent the equations of motion in the
form~\eqref{eq_11}, and, in this case, $K=K(\bold q)$ (i.e., $K$ is a
function depending on~$\bold q$ only).

In that case, it follows from~\eqref{eq_12} that ${\text\bf J}$ defines a
Poisson bracket, and the equations of motion are represented in
conformally Hamiltonian form.
\qed

\begin{rem}
If a generalized Chaplygin system has a singular invariant
measure with density $\mathcal{N}(\bold q)$ (depending on the configuration
variables only), then the equations of motion are represented in
conformally Hamiltonian form, except for the domain in the phase space in
which the density of the invariant measure either has a singularity or
vanishes.
\end{rem}

Below we illustrate Theorem~1 and the above arguments by considering
several problems of nonholonomic mechanics.

\section{Motion of a mass point}

Consider the motion of a mass point in Euclidean space $\Bbb R^3$. In this
case, $(q_1,q_2,q_3)$ are the Cartesian coordinates of the point, and the
Lagrangian function is of the form
\begin{equation}
\label{eq_14}
L=\frac12m(\dot q_1^2+\dot q_2^2+\dot q_3^2)-U(\bold q), \quad \bold
q=(q_1,q_2),
\end{equation}
where $m$ is the mass of the particle and $U(\bold q)$ is the potential of
the external forces. In this case, condition \eqref{eq_9} can be represented
as
\begin{equation}
\label{eq_15}
\frac\partial{\partial q_1}\left(\frac{a_1}{1+a_1^2+a_2^2}\left(\frac{\partial
a_1}{\partial q_2}- \frac{\partial a_2}{\partial
q_1}\right)\right)+\frac\partial{\partial
q_2}\left(\frac{a_2}{1+a_1^2+a_2^2}\left(\frac{\partial a_1}{\partial q_2}-
\frac{\partial a_2}{\partial q_1}\right)\right)=0.
\end{equation}
Let us consider three examples in more detail.

\subsection{Nonholonomic Oscillator}

Let the nonholonomic constraint be of the form
\begin{equation}
\label{eq_16}
\dot q_3-q_2\dot q_1=0.
\end{equation}

In accordance with~\cite{2}, we refer to the system thus obtained as a
{\it nonholonomic oscillator}. It is usually associated with the book of
Rosenberg~\cite{34}, although it was considered much earlier by Bottema
\cite{18} from the viewpoint of equilibrium positions and their stability
and by Hamel~\cite{25} (the potential field of this system is usually
assumed to be quadratic in $\bold q$).

\begin{rem}
Note that in~\cite{43} an implementation of the constraint
\eqref{eq_16} using a~plate sliding on a knife edge was suggested. Here
$(q_2,q_3)$ are the coordinates of the center of mass of the plate and
$q_1$ is the rotation angle of the plate (i.e., the configuration space in
this case is $\Bbb R^2\times S^1$).
\end{rem}

In the case under consideration, $a_1(\bold q)=q_2$ and $a_2(\bold q)=0$; then
condition~\eqref{eq_15} holds, and the density of the invariant measure is
$$
\mathcal{N}(\bold q)=(1+q_2^2)^{\frac12}.
$$

Hence, after rescaling time $d\tau=\mathcal{N}dt$ (as follows from Theorem~1),
the equations of motion of the nonholonomic oscillator are represented in
Hamiltonian form with the canonical Poisson bracket and the Hamiltonian
\begin{equation}
\label{eq_17}
\overline H=\frac{p_1^2}{2m}+\frac{(1+q_2^2)p_2^2}{2m}+U(\bold q).
\end{equation}

For the case where there is no potential, this result was obtained
in~\cite{35} and for the case $U(\bold q)={q_2^2}/2$ it was presented
in~\cite{23}.

Moreover, it turns out that, after the canonical change of variables
$$
q_1=x, \qquad p_1=p_x, \qquad q_2=\ln((1+y^2)^{\frac12}+y), \qquad
p_2=(1+y^2)^{\frac12}p_y,
$$
the Hamiltonian $\overline H$ becomes
\begin{equation}
\label{eq_18}
H^{(p)}=\frac{p_x^2+p_y^2}{2m}+V(x,y),
\end{equation}
where $V(x,y)=U(q_1,q_2(y))$.\goodbreak

Thus, the problem reduces to the investigation of the motion of a mass
point with (Cartesian) coordinates ($x$, $y$) on a plane, in the potential
field $V(x,y)$. The isomorphisms found (for a nonholonomic oscillator and
the Heisenberg system) provide a natural explanation of the possibility of
adding integrable potentials presented in~\cite{37a}.

\subsection{Heisenberg System}

Consider another example, namely, the {\it Heisenberg system}, for which
the\linebreak nonholonomic constraint is represented as
\begin{equation}
\label{eq_19}
\dot q_3-q_2\dot q_1+q_1\dot q_2=0,
\end{equation}
and which apparently first appeared in the book~\cite{20} in connection
with control problems. Other nonholonomic systems (involving rolling
motion) were\linebreak considered in~\cite{11, 27, 39} in connection with control
problems.

The authors of \cite{32} consider the motion of a point in a potential
field of the form
$$
U(\bold q)=\frac12(\alpha_1q_1^2+\alpha_2q_2^2)
$$
and prove, using the Poincar\'e mapping, that in this case the system
exhibits chaotic behavior. It can be seen from the Poincar\'e map that the
behavior of the trajectories of this system is similar to that of the
trajectories of nonintegrable two-degree-of-freedom Hamiltonian systems.

It turns out that this similarity is not accidental and is due to the fact
that in this case Theorem~1 applies for an (arbitrary) potential field
$U(\bold q)$.

Indeed, in this case, condition \eqref{eq_15} holds identically, and the density
of the invariant measure is of the form
$$
\mathcal{N}(\bold q)=(q_1^2+q_2^2+1)^{-1}.
$$
Thus, the equations of motion are represented in conformally Hamiltonian
form with Hamiltonian
\begin{equation}
\label{eq_20}
 \overline
H=\frac{1+q_1^2+q_2^2}{2m}((q_1p_1+q_2p_2)^2+p_1^2+p_2^2)+U(\bold q)
\end{equation}
and a canonical Poisson bracket.

It turns out that, in this case, the system \eqref{eq_20} reduces to the
investigation of the motion of a mass point on the sphere $S^2$ in a
potential field. Indeed, let us carry out the central projection
(for details, see, e.g., \cite{4}),
$$
q_1=\tan\theta\cos\varphi, \qquad q_2=\tan\theta\sin\varphi,
$$
where $\theta\in(0,\pi)$ and $\varphi=[0,2\pi)$, and pass to the
(canonically conjugate) momenta
$$
p_\theta=-\frac{1+q_1^2+q_2^2}{\sqrt{q_1^2+q_2^2}}(q_1p_1+q_2p_2), \quad
p_\varphi=q_2p_1-q_1p_2.
$$
As a result, the Hamiltonian~\eqref{eq_20} becomes
$$
H^{(s)}=\frac{p_\theta^2}{2m}+\frac{p_\varphi^2}{2m\sin^2\theta}+V(\theta,\varphi),
$$
where $V(\theta,\varphi)=U(q_1(\theta,\varphi),q_2(\theta,\varphi))$.

Because of the ambiguity of the central projection (taking two points on
a~sphere into one on a~plane), the above isomorphism is defined only on
half the hemisphere, and all the trajectories crossing the equator
($\theta=\frac\pi2$) are transformed into infinite trajectories on the
plane.

\subsection{Blackall Nonholonomic Constraint}

In conclusion of the present section, consider the constraint $ \dot
q_3-q_1q_2\dot q_1=0 $ (suggested in~\cite{5}), for which condition
\eqref{eq_15} does not hold. In other words, in this case there is no
invariant measure with density $\mathcal{N}(\bold q)$ (depending on the
configuration variables only), and hence, Theorem~1 does not apply.

In this case, the equations of motion \eqref{eq_6} become
\begin{equation}
\label{eq_21}
\begin{aligned}
\dot P_1&=\frac{q_1q_2P_1}{m(1 + q_1^2q_2^2)}
\left(\frac{q_2P_1}{1 + {q_1}^2{q_2}^2} - q_1P_2\right)\!, \quad
\dot P_2=\frac{2q_1^2q_2P_1^2}{m(1 + q_1^2q_2^2)^2}\\
\dot q_1&=\frac{P_1}{m(1+q_1^2q_2^2)}, \quad \dot q_2=\frac{P_2}m.
\end{aligned}
\end{equation}
It turns out that system \eqref{eq_21} has a singular invariant measure
(depending on phase variables)
$$
(1+q_1^2q_2^2)^{\frac14}|P_1|^{\frac12}d\bold qd\bold P.
$$
Note that system \eqref{eq_21} has the following family of particular solutions:
\begin{equation}
\label{eq_22}
P_1=0, \qquad P_2=\text{\rm const}, \qquad q_1=\text{\rm const}, \qquad
q_2=\frac{P_2}mt,
\end{equation}
in which the density of the invariant measure has a singularity.

Numerical experiments show that system \eqref{eq_21} exhibits asymptotic
behavior on the time interval $t\in(-\infty,+\infty)$, i.e., as
$t\to-\infty$, the motion begins with an unstable solution~\eqref{eq_22}
and, as $t\to+\infty$, tends to a stable solution. Thus,
system~\eqref{eq_21} exhibits a behavior that differs substantially from
that in the Hamiltonian case; see also \cite{31} and~\cite{33}.

\section{Chaplygin sleigh on a plane}

In this section, we consider the motion of a Chaplygin sleigh on a
horizontal fixed plane. As a rule, by the Chaplygin sleigh one means a
rigid body in the plane supported at two (or more) absolutely smooth legs
and a sharp weightless wheel (disk or knife edge), which prevents its
contact point $P$ from slipping in the direction perpendicular to the
plane of the wheel (see~Fig.~1).

\begin{figure}[!ht]
\centering
\includegraphics{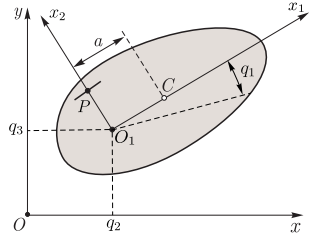}
\caption{Chaplygin sleigh on a plane}
\end{figure}

Note that, as an example illustrating the reducing multiplier
method,\linebreak Chaplygin \cite{46} considered the motion of the
Chaplygin sleigh. However, his considerations use substantially the
quasi-coordinate introduced by him (which gave rise to a debate concerning
the correctness of the method~\cite{44}).

\begin{rem}
It is of interest that the reducing multiplier method has
allowed the equations of motion to be represented in conformally
Hamiltonian form in another well-known problem, due to Chaplygin
\cite{45}, on the rolling motion of a~dynamically asymmetric ball on a
horizontal plane (for various generalizations of this problem,
see~\cite{3, 6, 12, 14}). A qualitative and topological analysis of the
motion of the contact point of the Chaplygin ball has been made recently
in~\cite{10}.
\end{rem}

In what follows we prove that the reducing multiplier method is applicable
only in the case $a=0$, i.e., if the~center of mass, $C$, is placed on the
perpendicular to the plane of the knife edge passing through the contact
point $P$.

As a historical remark, we note that, although the Chaplygin sleigh
is\linebreak customarily associated with the works of Chaplygin~\cite{46}
and Carath\'eodory~\cite{21}, it was considered somewhat earlier by Brill
in the book~\cite{19}, as an example of the mechanism of a nonholonomic
planimeter.

\begin{rem}
Diverse generalizations (variations) of the problem of
motion of the Chaplygin sleigh were considered in~\cite{41, 16}. For
example, in \cite{41}, the motion of the Chaplygin sleigh with torque and
on an inclined plane in a gravitational field was considered, and in
\cite{16} the equations of motion were obtained and the equilibrium
positions were studied for the Chaplygin sleigh on a rotating plane.
\end{rem}

Introduce two coordinate systems: an inertial (fixed) one, $Oxy$,
and\linebreak a~noninertial coordinate system $O_1x_1x_2$ attached to the
Chaplygin sleigh\linebreak (see~Fig.~1).

The configuration space in this case coincides with the motion group of
the plane $SE(2)$. To parameterize this space, we choose the angle $q_1$
of rotation of the axes of~$Oxy$ with respect to $Ox_1x_2$ and the
Cartesian coordinates $(q_2,q_3)$ of point $O_1$ in the coordinate system
$Oxy$. Then the constraint equation in the chosen variables can be
represented as
$$
\dot q_3-\frac{\cos q_1}{\sin q_1}\dot{q_2}=0.
$$
The Lagrangian function is
$$
L=\frac12m(\dot q_2^2+\dot q_3^2)+\frac12Iq_1^2+ma\dot q_1(-\dot q_2\sin
q_1+\dot q_3\cos q_1)-U(\bold q),
$$
where $U(\bold q)$ is the potential, and $m$ and $I$ are, respectively,
the mass of the body and its moment of inertia relative to point~$O_1$. As
a result, we obtain
$$
S=\frac{\cos q_1}{\sin q_1}\left(\frac{2maP_1\sin
q_1-ma^2P_2\cos2q_1+IP_2}{I-ma^2\cos^22q_1}\right)\!.
$$
A straightforward verification shows that relation~\eqref{eq_10} holds only
for $a=0$. In this case, we find a singular invariant measure (a reducing
multiplier) in the form $ \mathcal{N}=\sin q_1$.

Note that, for $a\ne0$, the dynamics of the system is of asymptotic
nature, which is why there is no (smooth) invariant measure with density
of the form~$\mathcal{N}(\bold q)$.

\begin{rem}
Nevertheless, for $U(\bold q)=0$, but $a\ne0$, one can
represent the equations of motion in Hamiltonian form~\cite{41} using the
method developed in~\cite{26}.
\end{rem}

Applying Theorem~1 (for $a=0$), we find the Hamiltonian
$$
\overline H=\frac{p_1^2}{2I\sin^2q_1}+\frac{p_2^2}{2m}+U(\bold q),
$$
which, after the change of variables
$$
x=\sqrt{\frac Im}\cos q_1, \quad p_x=\sqrt{\frac mI}\frac{p_1}{\sin q_1}, \quad
y=q_2, \quad p_y=p_2,
$$
reduces to the Hamiltonian \eqref{eq_18}, i.e., in this case, the problem
reduces to investigating the motion of a mass point on a plane.

\begin{rem}
In this example, in the constraint equation the singularity\linebreak at
${q_1=0,\pi}$ is related to the choice of local coordinates. For this
reason, in some cases, it is more convenient to study the equations of
motion in quasi-coordinates.
\end{rem}

\section{Conclusion}

In conclusion, we discuss some open problems.

Above we have obtained equation \eqref{eq_15} for Euclidean metric $\Bbb
R^3$; this equation describes nonholonomic constraints of Chaplygin type
for which the system admits a conformally Hamiltonian representation.
Moreover, for the two examples of constraints, \eqref{eq_16}
and~\eqref{eq_19}, in the absence of an external field, the system turns out
to be integrable. In this connection, it would be of interest to find a
general parameterization of constraints satisfying \eqref{eq_15} and to find
out in the general case whether  the system is integrable without
potential. We also note that, for holonomic systems, there are projective
transformations \cite{1} preserving the trajectories of the metrics on the
plane and on the sphere. The existence of such transformations for the
systems considered remains an open problem.

As was shown above, by combining different constraints and potentials, one
can obtain diverse systems that exhibit various effects typical of
nonholonomic mechanics. It would be of interest to choose the simplest of
these systems, which (for different parameter values) would possess all
unusual properties of nonholonomic systems, such as limit cycles, strange
attractors, etc. (a similar problem statement was considered
in~\cite{38}).

\end{document}